%% file: bellalta_WCL2022-1899-R1_main.tex
\documentclass[lettersize,journal]{IEEEtran}
\usepackage{comment}
\usepackage[table]{xcolor}

\usepackage{booktabs} % For formal tables
\usepackage{cite}
\usepackage{amsmath,amssymb,amsfonts}
\usepackage{graphicx}
\usepackage{xspace}
\usepackage{textcomp}
\usepackage{subcaption}
\usepackage{booktabs}
\usepackage{multirow}
\usepackage{subfiles}
\usepackage{xspace}
\usepackage{enumitem}

\usepackage{mathtools}
\usepackage{soul}

\newlength \figwidth
\newcommand{\red}[1]{{\textcolor[rgb]{1,0,0}{#1}}}

\definecolor{boristext}{rgb}{0.00, 0.00, 0.00}
\definecolor{boriscomments}{rgb}{0.88, 0.04, 0.04}
\newcommand{\bb}[1]{\textcolor{boristext}{#1}} % for text added
\newcommand{\tb}[1]{\textcolor{blue}{#1}}

\DeclareMathOperator{\E}{\mathbb{E}}

\begin{document}

\title{Delay Analysis of IEEE 802.11be\\Multi-link Operation under Finite Load}

\author{
\IEEEauthorblockN{Boris Bellalta, Marc Carrascosa, Lorenzo Galati-Giordano, and Giovanni Geraci
}
\thanks{B. Bellalta, M. Carrascosa, and G. Geraci are with Univ. Pompeu Fabra, Barcelona, Spain. They were supported in part by grants PID2021-123995NB-I00, PGC2018-099959-B-I00, PRE2019-088690,
RTI2018-101040-A-I00, PID2021-123999OB-I00, and by the ``Ram\'{o}n y Cajal" program.}
\thanks{L. Galati-Giordano is with Nokia Bell Labs, Stuttgart, Germany.}
}

\maketitle

\input{00_Abstract}

\begin{IEEEkeywords}
Multi-link Operation, IEEE 802.11be, Wi-Fi 7.
%delay, unlicensed spectrum, WLAN.
\end{IEEEkeywords}

\input{01_Introduction}
\input{02_SystemModel}
\input{03_Analysis}
\input{04_NumericalResults}
\input{05_Conclusion}

\bibliographystyle{IEEEtran}
\bibliography{Refs}

\end{document}

%% file: 00_Abstract.tex
\begin{abstract}
Multi-link Operation (MLO), arguably the most disruptive feature introduced in IEEE 802.11be, will cater for delay-sensitive applications by using multiple radio interfaces concurrently. In this paper, we analyze the delay distribution of MLO under non-saturated traffic. Our results show that upgrading from legacy single-link to MLO successfully contains the $95$th percentile delay as the traffic load increases, but also that adding extra interfaces yields rapidly diminishing gains. 
Further experiments we conduct on Google Stadia traces, accounting for realistic traffic with batch arrivals, confirm the insights stemming from our theoretical analysis, vouching for MLO to support
high-resolution real-time video with a controlled delay.
\end{abstract}

%% file: 01_Introduction.tex
\section{Introduction}

Wi-Fi is more popular than ever, with 628 million Wi-Fi hotspots by 2023, four times up from 2018, 11\% of which adopting Wi-Fi 6 or 6E \cite{garcia2021ieee}. Meanwhile, a new generation of Wi-Fi---IEEE 802.11be, or Wi-Fi 7---is in the making, with technical discussions underway to determine the specific implementation of several disruptive new features~ \cite{khorov2020current,lopez2019ieee,deng2020ieee,yang2020survey}. The new capabilities of IEEE 802.11be may include 320~MHz channels, 16 spatial streams, and multi-link operation (MLO)~\cite{CarGerGal2022}. MLO propounds the joint use of multiple radio interfaces on a single device through a single association. Owing to its promised performance gains, MLO is arguably the new feature drawing the most attention from industry and academia alike. 

Relevant contributions exist---including both analytical and simulation results---on the achievable throughput for different MLO implementations under full-buffer conditions \cite{song2020performance,korolev2022analytical,jin2022performance}, on the impact of MLO when coupled with access point (AP) coordination \cite{yang2019ap}, and on MLO traffic allocation policies \cite{lopez2022multi, lopez2022dynamic, suer2022adaptive}. As for the delay, the impact of MLO was studied in dense and demanding scenarios \cite{lacalle2021analysis,carrascosa2021experimental} and also in the presence of legacy single-link operation (SLO) devices \cite{naik2021can,murti2022multilink,carrascosa2022performance}.

In this paper, we carry out an analytical study of the delay for MLO under finite load. Different from works considering saturated traffic \cite{song2020performance,korolev2022analytical,jin2022performance}, finite-load models capture the interplay between the traffic, buffer, and network dynamics, hence allowing one to also analyze delay besides throughput. Understanding whether and how MLO is able to guarantee a certain worst-case delay is nowadays one of the most compelling challenges for the Wi-Fi research community, and thus the main motivation of this work.

Out of the possible implementations currently under consideration by the IEEE 802.11be Task Group, we focus on Simultaneous Transmit and Receive Multi-Link Multi-Radio (STR MLMR). STR MLMR allows running multiple backoff instances per packet, transmitting the packet on the interface whose backoff ends first. It is thus the implementation potentially achieving the highest gains in terms of throughput increase and delay reduction \cite{carrascosa2022performance}.
Our main findings are:
\begin{itemize}[leftmargin=*]
    \item
	By using multiple interfaces, MLO controls the $95$th percentile delay as traffic load increases. Nevertheless, adding extra interfaces yields diminishing gains, as the probability that an incoming packet gets buffered rapidly decays.
	\item
    Given a fixed tolerable delay, scaling up the number of MLO interfaces by a certain factor yields an even higher increment in the traffic load supported. Such gains escalate as the channel occupancy grows, i.e., in crowded scenarios.
    \item
	Experiments conducted with Google Stadia cloud gaming traces---accounting for real traffic with batch arrivals---confirm our theoretical analysis, vouching for MLO to support increasing video resolutions with a controlled delay.
%	\bb{Simulations using} Google Stadia cloud gaming traces---accounting for real traffic with batch arrivals---confirm our theoretical analysis, vouching for MLO to support increasing video resolutions with a controlled delay.
\end{itemize}

%% file: 02_SystemModel.tex
\section{System Model}
\label{Sec:Model}

In this section, we introduce STR MLMR and the modeling assumptions we make to characterize its delay performance.

\subsection{\!Simultaneous Transmit and Receive Multi-Link Multi-Radio}

\begin{figure*}
    \centering
    \includegraphics[width=0.80\textwidth]{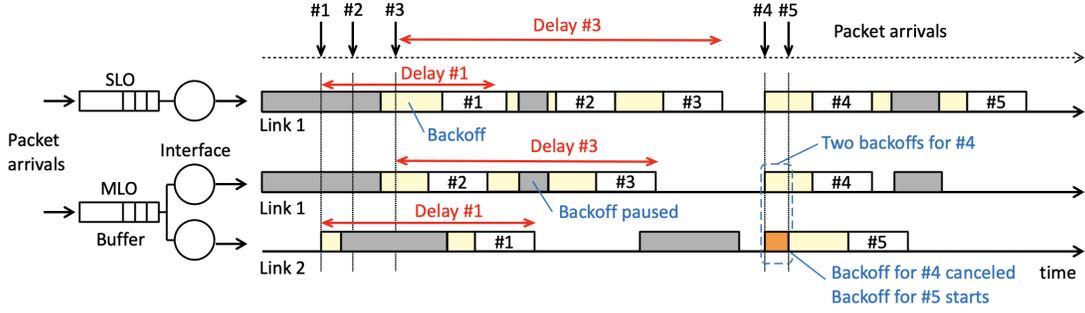}
    \caption{Exemplified illustration of SLO (top) and MLO (bottom) under its STR MLMR implementation. Grey, yellow/orange, and white slots denote occupied channels, ongoing/canceled backoffs, and successful transmissions, respectively.}
    \label{Fig:MLO_howitworks_b}
\end{figure*}

In STR MLMR, all interfaces are operated independently and can access different links asynchronously, either in the same or in different frequency bands \cite{CheCheDas22}.
For instance, STR MLMR with four links, each on an 80~MHz channel, could be implemented by using two channels each in the 5~GHz and 6~GHz bands, with a minimum channel separation of 160~MHz, and equipping devices with suitable radio-frequency filters \cite{CarGerGal2022}. While STR MLMR relies on self-interference suppression, it provides greater flexibility when compared to alternative MLO implementations \cite{carrascosa2022performance}.

At the transmitter side, when the number of packets in the system is lower than the number of interfaces, an arriving packet can benefit from multiple backoff instances, one on each inactive radio interface. The arriving packet waits in the buffer until it is allocated to an interface whenever: $i$) one of its backoff instances expires, or $ii$) another packet arrives, whereby the previous packet is allocated to the interface with the closest-to-expire backoff, and a new backoff is initiated on the other interface(s) for the new packet. Fig.~\ref{Fig:MLO_howitworks_b} illustrates the mode of operation of STR MLMR (bottom)---where both available radio interfaces, represented as circles, share a single buffer---alongside SLO (top), used as a benchmark and where packets are transmitted sequentially. \bb{From Fig.~\ref{Fig:MLO_howitworks_b}, we observe that:}

%At the transmitter side, a backoff instance is initiated upon packet arrival on all inactive radio interfaces. Incoming packets wait in the buffer until they are allocated to an interface whenever: $i$) one of the backoff instances expires, or $ii$) a new packet arrives, whereby the previous packet is allocated to the interface with the closest-to-expire backoff, and a new backoff is initiated on the other interface(s) for the new packet. Fig.~\ref{Fig:MLO_howitworks_b} illustrates the mode of operation of STR MLMR (bottom)---where both available radio interfaces, represented as circles, share a single buffer---alongside SLO (top), used as a benchmark and where packets are transmitted sequentially. \bb{From Fig.~\ref{Fig:MLO_howitworks_b}, we observe that:}

\begin{itemize}[leftmargin=*]
%
%\item \bb{When Packet \#1 arrives, it is allocated to Link 2 because Link~1 is busy.} 
\item When Packet \#1 arrives, although Link~1 is busy, both links are available, thus it contends on both links.
%
%\item When Packet \#2 arrives, Link~1 is still occupied while Link~2 is being used for Packet \#1. Packet \#2 is allocated to Link~1 since this is the first link to become available. 
\item When Packet \#2 arrives, since Link~1 is still busy, Packet \#1 is allocated to Link~2 as it has the closest-to-expire backoff, and Packet \#2 is allocated to Link~1. 
\item \bb{When Packet~\#3 arrives, Links~1 and 2 are occupied. Packet~\#3 waits in the buffer until Link 1 becomes available.}
\item Packet \#4 arrives when both links are available, thus two new backoffs are initiated on both Link~1 and Link~2.
\item \bb{When Packet \#5 arrives, the closest-to-expire backoff for Packet \#4 is maintained (on Link 1) while the other (on Link 2) is canceled as Link 2 is allocated to Packet \#5}.
\item The MLO AP is then allowed to start a new backoff for Packet \#5 on Link 2. Packets \#4 and \#5 are transmitted once their respective backoffs on Link 1 and Link 2 expire.
\end{itemize}

% ---------------------------------------------------------
% ---------------------------------------------------------
% ---------------------------------------------------------
% ---------------------------------------------------------

\subsection{System Setup}

\subsubsection*{Network topology}
%\st{We consider a WLAN scenario comprising multiple Overlapping Basic Service Sets (OBSS)}. 
We consider an Overlapping Basic Service Sets (OBSS) WLAN scenario. For our analysis, we focus on a single BSS---denoted the \emph{target} BSS---consisting of an AP and a station (STA), both equipped with $S$ radio interfaces, each operating on a different channel.\footnote{Spacing between channels and self-interference suppression capabilities are assumed sufficient to allow simultaneous transmission and reception.}
% as shown in Fig.~\ref{Fig:MLO_howitworks_a}.
The target BSS observes OBSS activity, characterized through a per-interface channel occupancy $\rho$ and a collision probability $p$, statistically equal on all interfaces and modeled in Section~\ref{Sec:ServiceTime}.

\subsubsection*{Traffic model}
We only consider downlink traffic for the target BSS. We assume that traffic arrivals follow a Poisson process with rate $\lambda$ packets per second, and that the packet transmission duration is exponentially distributed with a mean of $\E[D_{s}]$ seconds. We assume the AP buffer to be infinite. In our numerical evaluations we assume a packet size of $L=12000$ bits and that all $S$ interfaces use the same transmission rate, with modulation and coding scheme (MCS) of 256-QAM 3/4, 2 spatial streams, and an 80 MHz channel bandwidth. The value of the minimum contention window (CW$_{\min}$) is set to 15 and the number of backoff stages ($m$) is set to $6$.

\subsubsection*{Backoff duration}
%Carrier Sense Multiple Access with Collision Avoidance (CSMA/CA) is used at each interface, with the respective instantaneous backoff values selected uniformly at random in $[0, \mathrm{CW}]$. 
In MLO, a given packet may be allocated $s \geq 1$ backoff instances (on different interfaces), selecting the one that ends first for transmission.
%with the shortest instantaneous backoff. 
We define the mean backoff duration $B$ (in slots) as the average value of such shortest backoff instance, conditioned on $s$ and in turn on the number of packets $n$ in the system. We derive the value of $B$ and its expectation $\E[B]$ over $n$ in Section~\ref{Sec:Backoff}.

\subsubsection*{Transmission duration}
Without considering the backoff, the duration of a successful packet transmission $T_{\mathrm{s}}$ and of a collision $T_{\mathrm{c}}$ can be respectively broken down as
\begin{equation}
\begin{aligned}
    T_{\mathrm{s}} & = T_{\rm rts} + \text{SIFS} + T_{\rm cts} + \text{SIFS} + T_{\rm data} \\
    &\quad+ \text{SIFS} + T_{\rm ack} + \text{DIFS} + \sigma,
\end{aligned}
\end{equation}
and
\begin{align}
    T_{\mathrm{c}}=T_{\rm rts} + \text{SIFS} + T_{\rm cts} + \text{DIFS} + \sigma,
\end{align}
where $T_{\rm rts}$, $T_{\rm cts}$, $\text{SIFS}$, $\text{DIFS}$, and $\sigma$ (duration of an empty backoff slot) are standard IEEE 802.11 parameters \cite{bianchi2000performance}. The remaining ones follow from the choice of MCS, number of spatial streams, and channel bandwidth.
\bb{For tractability, we assume all packets to be transmitted after a full backoff, thereby neglecting the post-backoff as well as those cases when incoming packets come across an idle interface and can be transmitted right after a DIFS.}\footnote{While generally accurate, this assumption loosens and yields conservative results for very low traffic load ($\lambda\rightarrow 0$) and OBSS activity ($\rho\rightarrow 0$).}

%% file: 03_Analysis.tex
\section{Delay Analysis}
\label{Sec:Analysis}

The MLO AP with $S$ interfaces can be seen as an M/M/S queueing system with packet arrival rate~$\lambda$. For each interface, the packet service rate $\mu$ and the traffic intensity $a$ are defined as 
\begin{equation}
\mu=\frac{1}{\E[D_{\mathrm{s}}]} \quad \text{and} \quad a=\frac{\lambda \E[D_{\mathrm{s}}]}{S},
\label{Eq:mu_a}
\end{equation}
where $\E[D_{\mathrm{s}}]$ is the expected service time, defined in the sequel.

\bb{Owing to the \textit{Poisson Arrivals See Random Arrivals} (PASTA) property, the probability that an incoming packet finds $n$ packets in the system is equal to the stationary probability $\pi_n$ \cite{wolff1982poisson}}, given by
\begin{equation}
    \begin{aligned}
	\pi_n=
	\begin{cases}
    \displaystyle
    \frac{(S a)^n}{n!}\pi_0, & \text{if $n < S$}\vspace*{10pt}\\
    \displaystyle
    \frac{S^S a^n}{S!}\pi_0, & \text{otherwise}
  \end{cases}
    \end{aligned}
    \label{eqn:pi_n}
\end{equation}
with
\begin{align}
    \pi_0 = \left(1+\sum_{n=1}^{S-1}{\frac{(S \, a)^n}{n!}}+\frac{(S \, a)^S}{S! \, (1-a)}\right)^{-1}\!\!\!\!\!\!\!.
    \label{eqn:pi_0}
\end{align}
Note that for $S=1$, the above M/M/S reduces to an M/M/1 system, hence modeling legacy SLO as a special case.

\subsection{Expected Service Time}
\label{Sec:ServiceTime}

The expected time required to successfully transmit a packet,~$\E[D_{\mathrm{s}}]$, can then be obtained as
\begin{align}
	\E[D_{\mathrm{s}}]=\frac{p}{1-p}\left(\frac{\E[B] \, \sigma}{1-\rho} + T_{\mathrm{c}} \right)+\left(\frac{\E[B] \, \sigma}{1-\rho} + T_{\mathrm{s}} \right), 
\label{Eq:service_time}
\end{align}
where $\E[B]$ is the expected value of the backoff duration per transmission, subsequently derived in (\ref{Eq:MeanBackoffOverN}), $\rho$ denotes the \emph{mean channel occupancy}, due to transmissions from other devices and affecting the backoff duration, and $p$ denotes the \emph{probability of collision}. In (\ref{Eq:service_time}), we assumed the latter to be the sole reason why a packet would not be received correctly, and that an unlimited number of retransmissions per packet is allowed, thus yielding a mean number of retransmissions $p/(1-p)$. 
The values of $\rho$ and $p$ in~\eqref{Eq:service_time} can be obtained from measurements data, simulations, or analytically as illustrated in Section \ref{Sec:Occupancy}.

% ---------------------------------------------------------
% ---------------------------------------------------------

\subsection{Expected Backoff Duration}
\label{Sec:Backoff}

%For legacy SLO (i.e., when $S=1$) the instantaneous backoff value is selected uniformly at random in $[0, \mathrm{CW}]$, resulting in a mean backoff duration of $B=\frac{\text{CW}}{2}$ slots.

%\cite{bellalta2019ap}
\bb{Following the use of binary exponential backoff with $m$ stages, the average backoff contention window in slots per transmission depends on the collision probability $p$ and is equal to \cite{bianchi2000performance}
\begin{align}
    \label{Eq:BEB}
    \overline{\text{CW}}=\frac{1-p-p(2p)^m}{1-2p}\left(\text{CW}_{\min}+1\right)-1.
\end{align}}
%\bb{Following the use of binary exponential backoff with $m$ stages, the average backoff duration in slots per transmission depends on the collision probability $p$ and is equal to \cite{bianchi2000performance}
%\begin{align}
%    \label{Eq:BEB}
%    \overline{\text{CW}}=\frac{1-p-p(2p)^m}{1-2p}\left(\frac{\text{CW}_{\min}+1}{2}\right)-\frac{1}{2}.
%\end{align}}
In the case of MLO, owing to its ability to allocate multiple backoff instances to a packet and to select the interface whose backoff expires first, the mean backoff duration per transmission can be computed as the minimum of $s$ i.i.d. uniform random variables, resulting in
\begin{align}
    B =\frac{\overline{\text{CW}}}{s+1},
\end{align}
where $s=\max{(1,S-n)}$ is the number of backoff instances allocated to a given packet on different links, all assumed to share the same statistical properties, and $n$ is the number of packets found when a packet is scheduled for transmission. 

%\st{in the system upon packet arrival}

Employing (\ref{Eq:service_time}) requires computing the expected backoff duration $\E[B]$ per transmission by marginalizing with respect to $n$, obtaining
\begin{equation} \label{Eq:MeanBackoffOverN}
    \E[B] = \sum_{n=0}^{S-1}{\pi_{n} \, \frac{\overline{\rm{CW}}}{S-n+1}}+\sum_{n=S}^{\infty}{\pi_{n}\frac{\overline{\rm CW}}{2}},
\end{equation}
where $\pi_n$ is given in (\ref{eqn:pi_n}).
Note that by increasing the traffic intensity $a$ in (\ref{eqn:pi_n}), the first summand in \eqref{Eq:MeanBackoffOverN} becomes progressively less dominant over the second one, and $\E[B]$ approaches the value $\frac{\overline{\rm CW}}{2}$ experienced for SLO. The latter confirms that it is under relatively low traffic loads when MLO best leverages the use of multiple interfaces to reduce the expected backoff duration.

\bb{Once obtained $\E[B]$, the transmission probability for the AP MLO on a given link is $\tau=\frac{\gamma}{\E[B]+1}$, with $\gamma=1-\sum_{n=0}^{S-1}{\frac{S-n}{S}\pi_n}$ the probability that a given interface is busy given that there are $n$ packets in the buffer.}

\subsubsection*{Remark}
The value of $\E[B]$ in \eqref{Eq:MeanBackoffOverN} depends on $\pi_{n}$ in \eqref{eqn:pi_n} and in turn on $\E[D_{\mathrm{s}}]$ and $\E[B]$ itself. Its computation therefore entails solving a fixed-point equation iteratively. %, as part of Algorithm~\ref{Algo}.

% ---------------------------------------------------------
% ---------------------------------------------------------

\bb{\subsection{Mean Channel Occupancy and Collision Probability}}
\label{Sec:Occupancy}

\bb{To illustrate how the values of $\rho$ and $p$ can be computed analytically, we consider the following simple case: the MLO AP is competing with $N$ contenders on each of its interfaces. We refer to the activity of the contenders as the OBSS activity. All contenders are using the same MCS and transmit packets of size $L$ as the MLO AP, and thus share the same value of $T_s$. Moreover, we define the activity factor of the contenders as $\alpha \in [0,1]$, with $\alpha=0$ and $\alpha=1$ respectively representing no activity and fully backlogged contenders.}

%\footnote{\bb{Note that $\alpha$ can be seen also as a control parameter to keep the system stable when contention increases.}} 

\bb{On each link where the AP is undergoing a backoff, any of the $N$ contenders can transmit with probability $\tau'$ at the beginning of each slot. Then, and following the definitions from~\cite{bianchi2000performance}, the MLO AP will observe an empty backoff slot with probability $p_{\mathrm{e}}=(1-\tau')^{N}$, a backoff slot containing a successful transmission with probability $p_{\mathrm{s}}=N \, \tau' \, (1-\tau')^{N-1}$, and a backoff slot containing a collision with probability $p_{\mathrm{c}}=1-p_{\mathrm{e}}-p_{\mathrm{s}}$.}

\bb{Under these conditions, the value of $\rho$ in \eqref{Eq:service_time} is obtained by noting that the expected backoff duration satisfies 
$\left(\frac{\E[B] \, \sigma}{1-\rho} \right) = \E[B] \, ( p_{\mathrm{e}} \, \sigma +  p_s \, T_{\mathrm{s}} + p_c \, T_{\mathrm{c}})$, which results in
\begin{equation}
	\rho = 1-\frac{\sigma  }{p_{\mathrm{e}} \, \sigma +  p_{\mathrm{s}} \, T_{\mathrm{s}} + p_{\mathrm{c}} \, T_{\mathrm{c}}}.
\label{Eq:rho}
\end{equation}
In addition, the collision probability $p$ in \eqref{Eq:service_time}, i.e., the probability that at least one of the $N$ contenders starts transmitting at the same time instant as the MLO AP, is given by
\begin{equation}
    p = 1-(1-\tau')^{N} = 1-p_{\mathrm{e}}.
\label{Eq:p}
\end{equation}}
\bb{Similar to the MLO AP, the transmission probability of a contending node, $\tau'=\frac{\alpha}{\overline{\rm \E[B']}+1}$, depends on the collision probability it observes, $p'=1-(1-\tau')^{N-1}(1-\tau)$, and thus on the MLO AP activity through $\tau$.}
\bb{\subsubsection*{Remark}
The values of $\tau$, $\tau'$, $p$, and $p'$ are mutually dependent. Their computation therefore entails solving a fixed-point equation iteratively.} %Moreover, note that at every iteration we must also solve the previously mentioned fixed-point equation to calculate $\E[B]$ and $\pi_n$, as described in Sec. \ref{Sec:Backoff}.

\begin{comment}
\subsubsection*{Example}
Consider a scenario with $N_{\mathrm{c}}=5$ contenders, $T_{\mathrm{c}} = 0.11313$~ms, $T_{\mathrm{s}} = 0.22127$~ms, a contention window $\text{CW}=15$, and the values listed in Section~\ref{Sec:Model} for packet size, MCS, number of spatial streams, and channel bandwidth. Under this setup, a low activity factor $\alpha=0.1$ corresponds to $p=0.057$ and $\rho=0.572$, whereas a higher activity factor of $\alpha=0.5$ yields $p=0.26$ and $\rho=0.85$. 
\end{comment}

%\subsubsection*{Example}
%Consider a scenario with $N=5$ contenders, $T_{\mathrm{c}} = 0.11313$~ms, $T_{\mathrm{s}} = 0.22127$~ms, and the values listed in Section~\ref{Sec:Model} for packet size, MCS, number of spatial streams, and channel bandwidth. Under this setup, given the MLO AP carries a traffic load of 10 Mbps, a low activity factor by the $N=5$ contenders of $\alpha=0.1$ corresponds to $p=0.053$ and $\rho=0.348$, whereas a higher activity factor of $\alpha=0.5$ yields $p=0.20$ and $\rho=0.76$.

% ---------------------------------------------------------
% ---------------------------------------------------------

\subsection{Delay Distribution}

We define the delay $D$ experienced by a packet as its queuing time in the buffer plus its service time $D_{\mathrm{s}}$. The delay $D$ is the response time of an M/M/S system which has an Erlang distribution given as follows \cite{jain1991art}
\begin{align}
	F_D(t)=
	\begin{cases}
    \displaystyle
    1 - e^{-\mu t} - \eta \, \mu \, t \, e^{-\mu t}
    & \text{if $a=\frac{S-1}{S}$}\vspace*{5pt}\\
    \displaystyle
    1-e^{-\mu t} - \eta \, \frac{e^{-S \mu (1-a) t }-e^{-\mu t}}{1 - S \, (1+a)}
    & \text{otherwise}
  \end{cases}
\label{Eq:CDF}
\end{align}
where
\begin{equation}
    \eta = \frac{(S \, a)^{S}}{S! \, (1-a)} \,\pi_0
\label{Eq:eta}
\end{equation}
is the probability that an incoming packet finds all $S$ interfaces busy and is placed in the buffer. 
Note that when $S=1$, (\ref{Eq:CDF}) reduces to the delay distribution of an M/M/1 queue, whereas for $\eta \rightarrow 0$ it boils down to the distribution of the service time~$D_{\mathrm{s}}$. 
%In Algorithm~\ref{Algo}, we detail the steps required to compute the delay distribution in (\ref{Eq:CDF}).

%\input{0X_Algorithm1}

%% file: 04_NumericalResults.tex
\section{Numerical Results} \label{Sec:Results}

In this section, we validate our analysis and characterize the delay performance of MLO. We further confirm our findings by comparing our theoretical results to those obtained under realistic non-Poisson traffic with batch arrivals.

% -----------------------------------------------------
% -----------------------------------------------------
% -----------------------------------------------------
% -----------------------------------------------------

\subsection{Theoretical Performance of MLO}\label{Sec:theoretical}

% -----------------------------------------------------
% -----------------------------------------------------

\subsubsection*{Analysis validation}

\begin{figure}
     \centering
     \begin{subfigure}[b]{0.155\textwidth}
         \centering
 	 	  \includegraphics[width=1\textwidth]{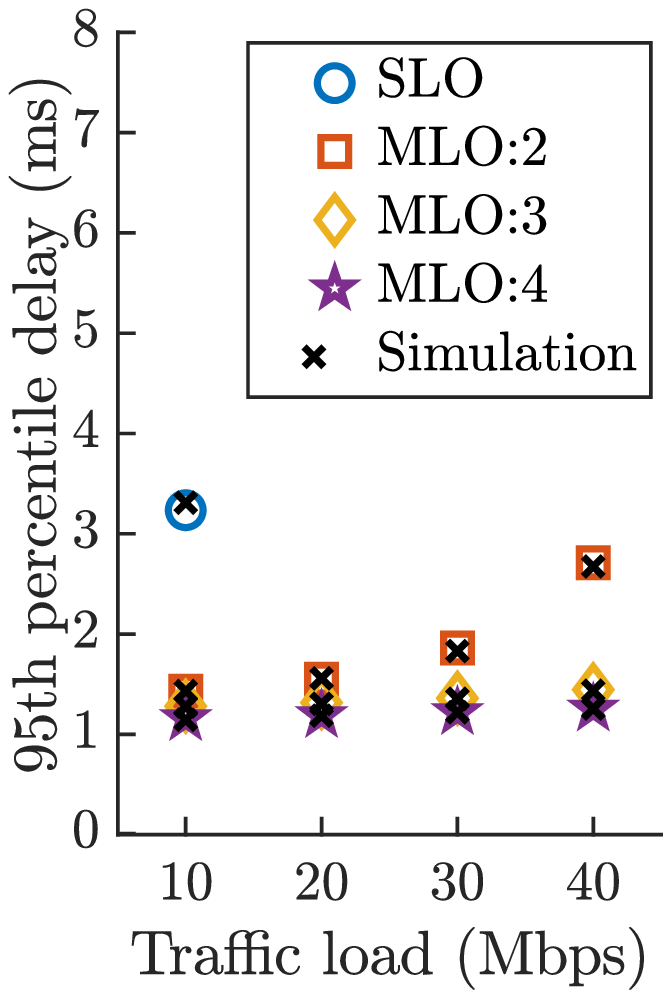}
 	 	  \caption{\bb{$N\!=\!5$; $\alpha\!=\!0.25$}}
 		  \label{Fig:Fig2a_BEB}
     \end{subfigure}
     \begin{subfigure}[b]{0.155\textwidth}
         \centering
 	 	  \includegraphics[width=1\textwidth]{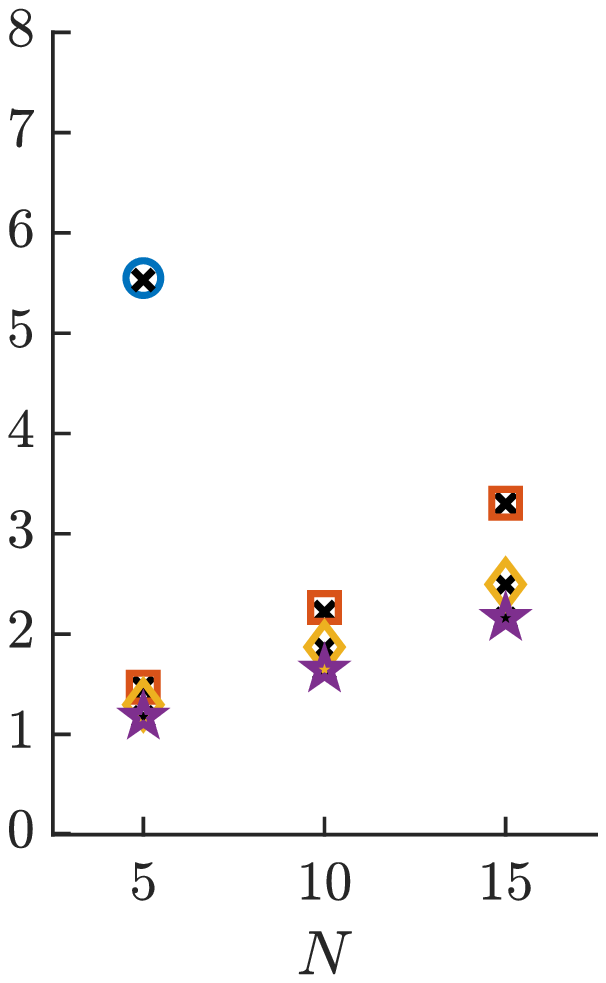}
 	 	  \caption{\bb{$15$ \!Mbps; $\alpha\!=\!0.25$}}
 		  \label{Fig:Fig2b_BEB}
     \end{subfigure}
     \begin{subfigure}[b]{0.155\textwidth}
         \centering
 	 	  \includegraphics[width=1\textwidth]{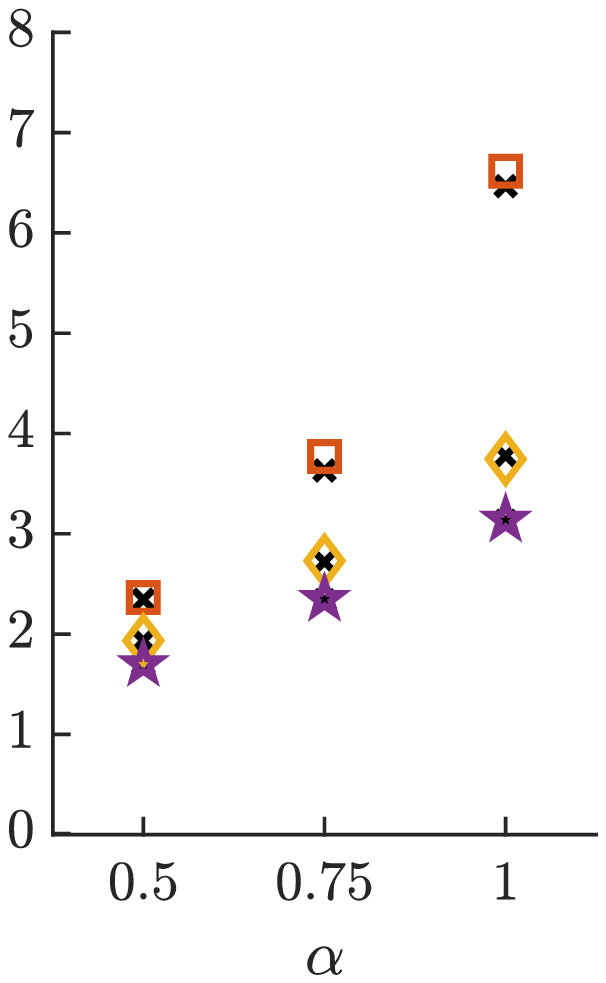}
                \caption{\bb{$10$ \!Mbps; $N\!=\!5$}}
         \label{Fig:Fig2c_BEB}
     \end{subfigure}
     \caption{\bb{95th percentile of the delay vs. (a) traffic load, (b) number of contenders, and (c) activity, under SLO and MLO with 2, 3, and 4 interfaces. Points are omitted for SLO when it incurs an unbounded delay. Crosses indicate values obtained via simulation for validation purposes.}}
     \label{Fig:CaseValidation}
\end{figure}

\bb{Fig.~\ref{Fig:CaseValidation} displays the 95th percentile of the delay---comprising queuing plus transmission time---for SLO and MLO with a variable number of interfaces vs. (a) the traffic load of the MLO AP when there are $N=5$ contenders with activity $\alpha=0.25$ (Fig.~\ref{Fig:Fig2a_BEB}); (b) the number of contenders $N$, when the load of the MLO AP is 15 Mbps and $\alpha=0.25$ (Fig.~\ref{Fig:Fig2b_BEB}); and (c) the activity $\alpha$, when $N=5$ and the load of the MLO AP is 10 Mbps (Fig.~\ref{Fig:Fig2c_BEB}).} Omitted circles indicate SLO inability support a certain load. Numerical values derived through our analysis (circles, squares, diamonds, and stars) are shown alongside those obtained via simulation (crosses), showing an excellent match.

% -----------------------------------------------------
% -----------------------------------------------------

\subsubsection*{Delay vs. number of interfaces}
Fig.~\ref{Fig:Fig2a_BEB} also shows that using multiple interfaces helps controlling the delay as the traffic load increases. For instance, given a 10~Mbps load, adding a second interface reduces the $95$th percentile delay by nearly three-fold. This behavior is exacerbated for higher loads, where MLO is the only way to guarantee a bounded delay. Nevertheless, adding more and more interfaces may result in anecdotal gains. Indeed, depending on the traffic load, the probability $\eta$ in (\ref{Eq:eta}) that an incoming packet is buffered vanishes by just adding one (or two) extra interface(s), with the packet delay thus driven exclusively by the service time.

\begin{comment}
\begin{figure}
    \centering
    \includegraphics[width=0.45\textwidth]{Figures/CaseValidation_Fig1.eps}
    \caption{95th percentile of the delay vs. traffic load under SLO and MLO with 2, 3, and 4 interfaces. Points are omitted for SLO when it incurs an unbounded delay. Crosses indicate values obtained via simulation for validation purposes. \textcolor{blue}{Add more info in the caption. This for 5 contenders, with $\alpha=0.1$}}
    %\bcom{Updated. $p = 0.05$ and $\rho = 0.57$. 10 Mbps = 720p; 30 Mbps = 1080 p; 40 Mbps = 2160p}
    \label{Fig:CaseValidation_Fig1}
\end{figure}
\end{comment}

% -----------------------------------------------------
% -----------------------------------------------------

\subsubsection*{Throughput vs. number of interfaces}

\begin{figure*}
     \centering
     \begin{subfigure}[b]{0.323\textwidth}
         \centering
 	 	  \includegraphics[width=0.9\textwidth]{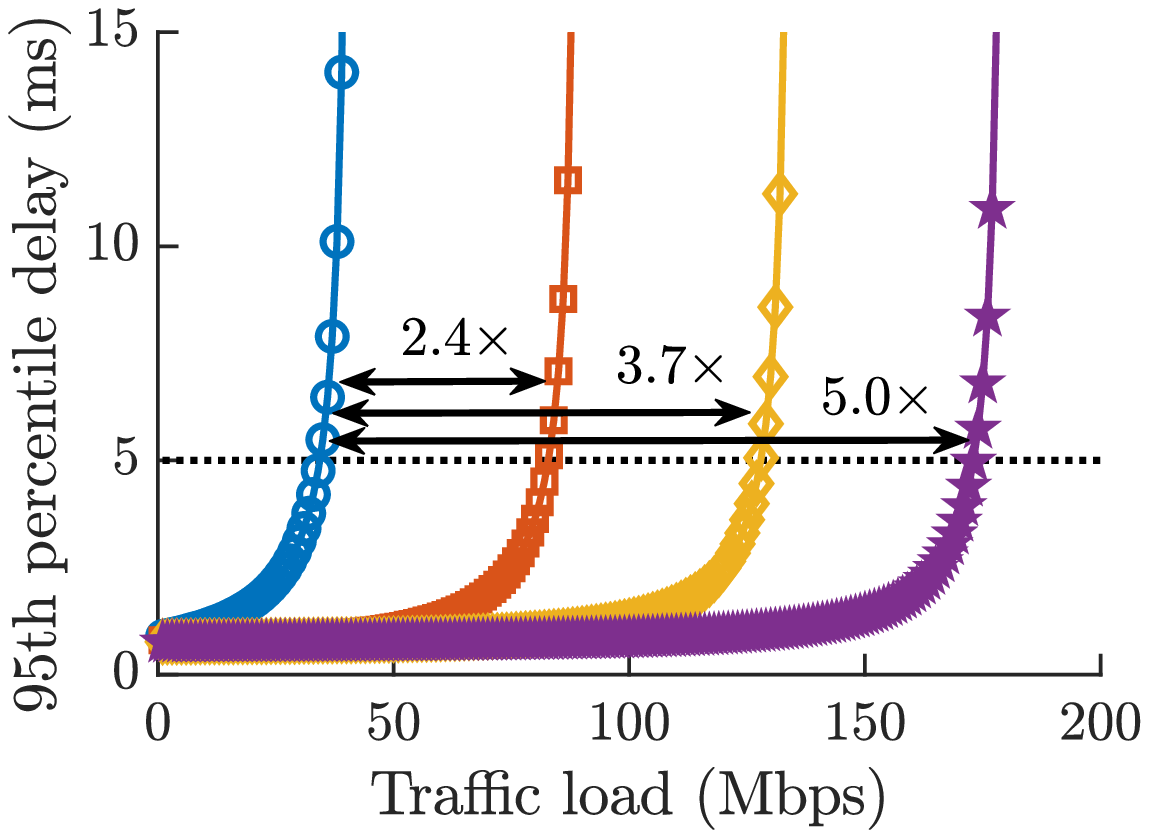}
 	 	  \caption{\bb{No OBSS activity ($\alpha=0$)}}\label{Fig:Fig3a_BEB}
     \end{subfigure}
     \hfill
     \begin{subfigure}[b]{0.323\textwidth}
         \centering
 	 	  \includegraphics[width=0.9\textwidth]{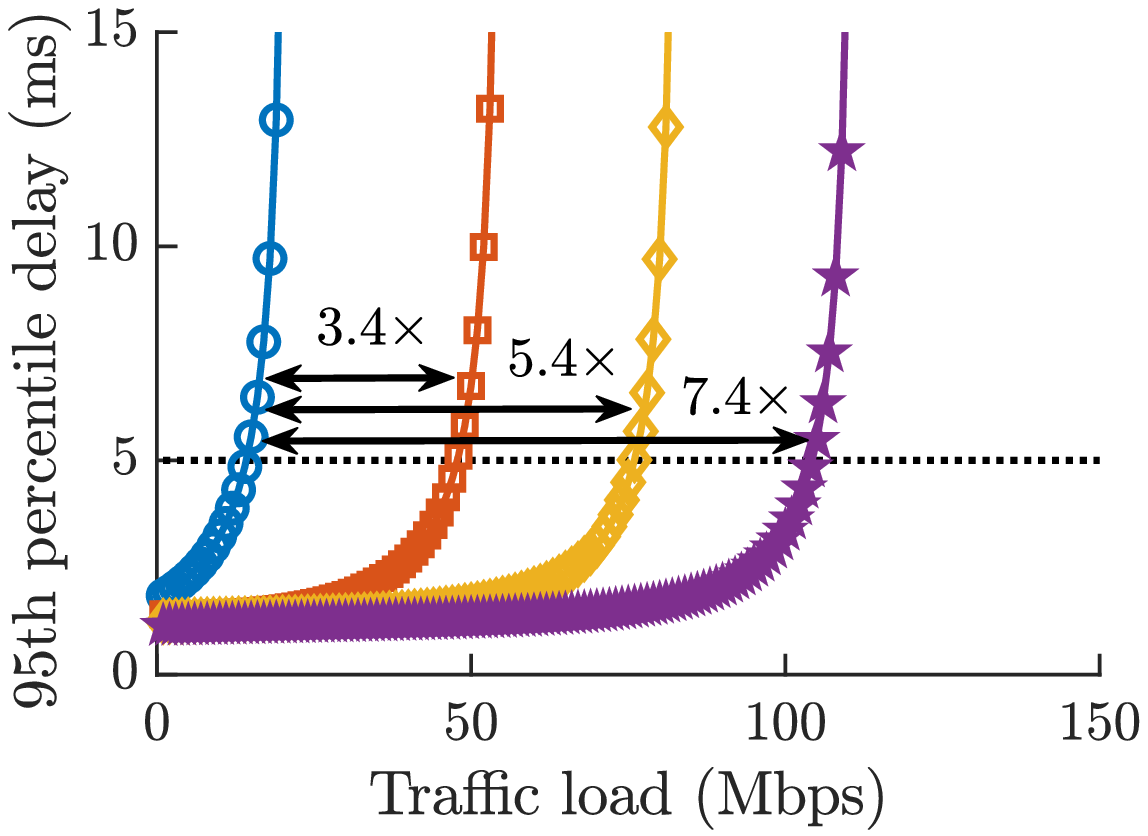}
         \caption{\bb{Med. OBSS activity ($\alpha=0.25$)}}
         \label{Fig:Fig3b_BEB}
     \end{subfigure}
     \hfill
     \begin{subfigure}[b]{0.323\textwidth}
         \centering
 	 	  \includegraphics[width=0.9\textwidth]{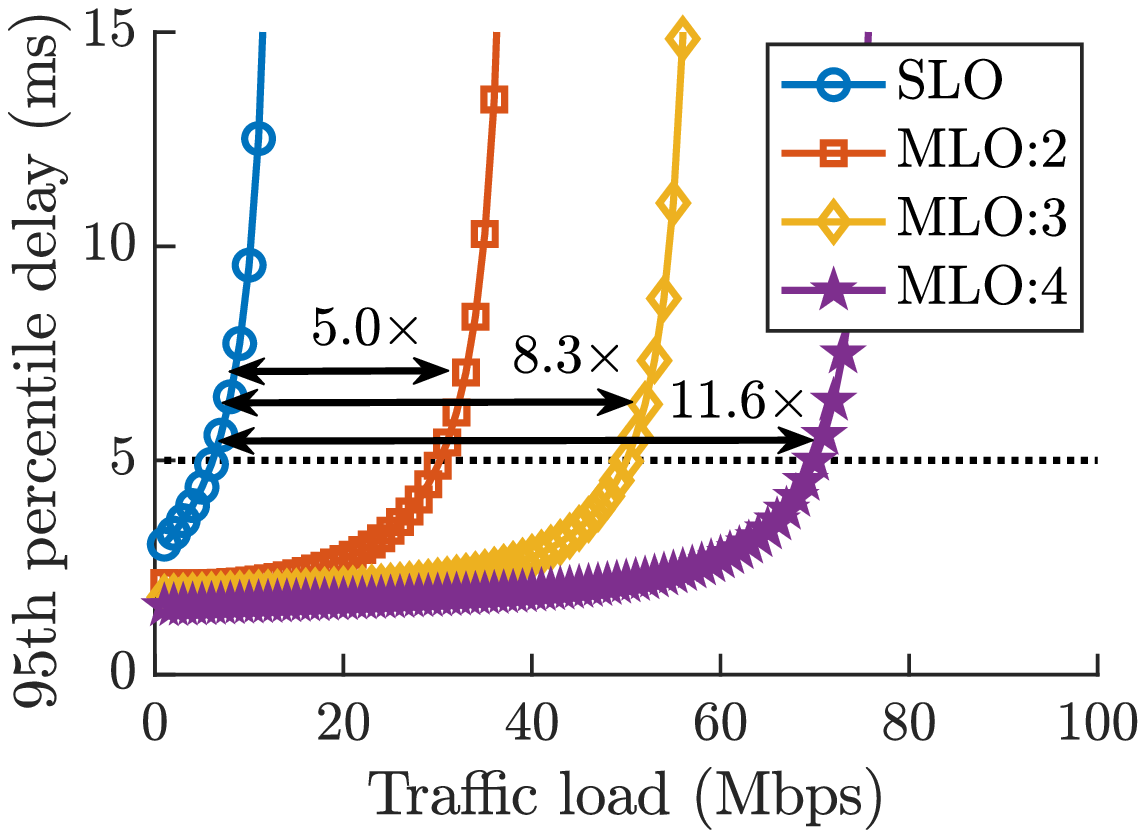}
         \caption{\bb{High OBSS activity ($\alpha=0.5$)}}
         \label{Fig:Fig3c_BEB}
     \end{subfigure}
     \caption{\bb{95th percentile of the delay vs. traffic load under SLO and MLO with a variable number of interfaces. No (a), medium (b), and high (c) OBSS activity is modeled by increasing the activity of the contenders.}}
     \label{Fig:CaseLoad}
\end{figure*}

The use of multiple interfaces sharing a single buffer notoriously yields multiplexing gains. Fig.~\ref{Fig:CaseLoad} shows the $95$th percentile of the delay vs. the traffic load in three different scenarios with increasing OBSS activity, allowing to compare the throughput---i.e., the traffic load supported---for a given allowed delay, for SLO and MLO under a variable number of interfaces. Specifically, we consider: (a) no contenders, (b) $N=5$ contenders with $\alpha=0.25$ (the same scenario as in Fig. \ref{Fig:Fig2a_BEB}) , and (c) $N=5$ with $\alpha=0.5$. \bb{We observe that under (a) no OBSS activity, fixing the $95$th percentile delay to $5$~ms results in $2.4\times$, $3.7\times$, and $5\times$ gains when adding a second, third, and fourth interface, respectively. Remarkably, such gains are higher than the actual scaling in the number of interfaces and they grow with the channel occupancy, exceeding $5\times$, $8\times$, and $11\times$ under (c) high OBSS activity, vouching for MLO as a tool to boost throughput in crowded scenarios.}

% -----------------------------------------------------
% -----------------------------------------------------
% -----------------------------------------------------
% -----------------------------------------------------

\subsection{Validation under Realistic Traffic} 
\label{Sec:Results2}

\begin{figure}[t]
    \centering
    \includegraphics[width=0.450\textwidth]{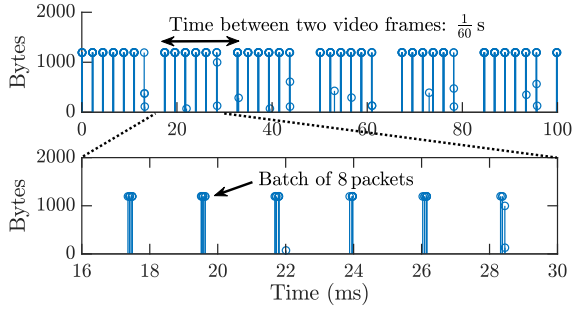}
    \caption{Google Stadia 100~ms traffic trace for 1080p video resolution (top) and zoom-in showing batch arrivals (bottom).}
    \label{Fig:StadiaCDFs}
\end{figure}

Since our theoretical results hinge on the assumptions of Poisson traffic and exponentially distributed service times, we now validate our main findings by considering realistic non-Poisson traffic with batch arrivals. The latter is a key feature to account for, being MLO capable of transmitting multiple packets in the same batch at once. We employ experimental traffic arrivals based on Google Stadia\footnote{Stadia is a cloud gaming service operated by Google. At the time of writing, Stadia lets users stream for free at resolutions of up to 1080p---and up to 4K under a paid subscription---at 60 frames per second, thus requiring high downlink throughput as well as consistently low delay.} traces \cite{carrascosa2022cloud}, replicating variable packet size and arrival patterns as exemplified in Fig.~\ref{Fig:StadiaCDFs}. We use the same simulator that validated our theoretical results in Section IV-A and maintain all parameters (other than the traffic generation) unchanged.

\begin{figure}[t]
    \centering
    \includegraphics[width=0.450\textwidth]{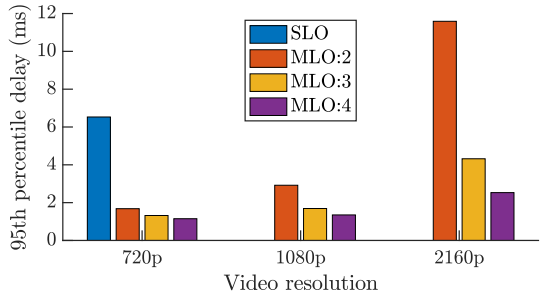}
    \caption{95th percentile delay vs. Google Stadia video resolution under SLO and MLO with variable number of interfaces. Bars are omitted for SLO when it incurs an unbounded delay.}
    %\bcom{Updated with: $p = 0.05$ and $\rho = 0.572$.}}
    \label{Fig:PlotStadiaVsPoisson2}
\end{figure}

Fig.~\ref{Fig:PlotStadiaVsPoisson2} reports the 95th percentile of the delay vs. Google Stadia video resolution under SLO and MLO with a variable number of interfaces, with omitted blue bars indicating the inability of SLO to support a certain resolution. Note that the resolutions $\{720\text{p}, 1080\text{p}, 2160\text{p}\}$ respectively correspond to the traffic loads of $\{10, 20, 40\}$~Mbps in Fig.~\ref{Fig:Fig2a_BEB}, where the same OBSS activity per link ($N=5$, $\alpha=0.25$) was also employed, thus allowing a direct comparison.
Similarly to what observed in Fig.~\ref{Fig:Fig2a_BEB} for Poisson traffic, SLO struggles to support increasing video resolutions (i.e., traffic loads) with a controlled delay. MLO circumvents this problem, proving able to successfully manage sporadic congestion episodes caused by large batch arrivals, although again, adding extra interfaces yields diminishing returns on the $95$th percentile delay.

%MLO circumvents this problem, although adding extra interfaces yields diminishing returns on the $95$th percentile delay.

%%%%%%%%%%%%%%%%%%%%%%%%%%%%%%%%%%
\begin{comment}
\tb{------------------------------------------------------------------------}
\red{Fig.~\ref{Fig:PlotTrafficArrivals} shows the impact that batch packet arrivals have on the 95th-percentile delay for both Poisson and CBR traffic under high load (25 Mbps) and medium channel occupancy ($\rho = 0.5$ and $p = 0.1$) conditions. In both cases, increasing the batch size (i.e., number of packets that arrive simultaneously) results in an increase in the worst-case delay. This increase is worse for Poisson traffic since multiple batches can arrive in a short period, causing a higher congestion level at the AP. This also explains why Poisson traffic benefits more from MLMR STR, showing gains of 1 order of magnitude (about a 10 ms reduction), while they are much more modest for CBR traffic (about a 2-3 ms reduction).}
\end{comment}

%% file: 05_Conclusion.tex
\section{Conclusion} \label{Sec:Conclusions}

In this paper, we introduced an analytical framework to study the delay distribution of MLO under non-saturated traffic, to advance the understanding of its potential benefits in the upcoming IEEE 802.11be, alas Wi-Fi 7. Among other things, our results showed how upgrading from legacy SLO to MLO successfully contains the $95$th percentile delay as the traffic load increases, but also that adding extra interfaces yields rapidly diminishing gains. %To account for realistic traffic with batch arrivals, we conducted further experiments on real Google Stadia traces. These confirmed the main insights stemming from our theoretical analysis, reaffirming MLO as a prime candidate to support high-resolution real-time video with a controlled delay in the unlicensed spectrum.

Our analytical framework is suitable for multiple extensions, ranging from a data-driven link-specific modeling of the channel occupancy and collision probability, to a characterization of features such as packet aggregation and dynamic channel bonding. 
\bb{Moreover, a study of its stability and its compatibility with legacy 802.11 backoff rules could offer new perspectives on the behavior of MLO.}
%\bb{Lastly, special attention should be placed on studying the stability of MLO throughput and delay regions, as well as the compatibility of new MLO backoff strategies with default IEEE 802.11 backoff rules.}

%Furthermore, while this work focused on STR MLMR, it may be desirable to capture other MLO modes analytically under finite load, allowing a direct comparison of their performance in different scenarios. 

\section*{Acknowledgments}
The constructive feedback from the Editor, Prof. Swades De, and the anonymous Reviewers is gratefully acknowledged.